\global\def\draftcontrol{0}

   \def\versionno{ superpotential }

\catcode`\@=11

\newcommand\makepapertitle{\par

  \begingroup
    \renewcommand\thefootnote{\@fnsymbol\c@footnote}%
 \newpage
     \global\@topnum\z@   
     \@makepapertitle
     \thispagestyle{empty}\@thanks
  \endgroup
  \setcounter{footnote}{0}%
  \global\let\thanks\relax
  \global\let\makepapertitle\relax
  \global\let\@makepapertitle\relax
  \global\let\@thanks\@empty
  \global\let\@author\@empty
  \global\let\@date\@empty
  \global\let\@title\@empty
  \global\let\title\relax
  \global\let\author\relax
  \global\let\date\relax
  \global\let\and\relax
  \def\version{\let\version\@version\@gobble}
}
\def\@makepapertitle{%
  \newpage
   \ifnum\draftcontrol=1 {}
   \version\versionno
   \vskip 5.5em%
   \else
   \hfill\hbox to 3.5cm {\parbox{5cm}{\@pubnum}\hss}%
   \vskip 6.5em%
   \fi
   \begin{center}%
   \let \footnote \thanks
      {\hskip -0\textwidth \hbox to 1\textwidth%
        {\centerline{\Large\bf{\noindent%
    \parbox[t]{1.3\textwidth}{\begin{center}\@title\end{center}}}}}}%
     \vskip 1.5em%
     {\normalsize
       \lineskip .5em%
       \begin{tabular}[t]{c}%
         \@author
       \end{tabular}\par}%
     \vskip 1.5em%
     {\@bstract}%
     \end{center}%
     \vfill
     \@date%
     \vskip 1.5em%
   \par
}

\gdef\@pubnum{}
\def\pubnum#1{%
  \gdef\@pubnum{#1}}

\gdef\@bstract{}
\def\Abstract#1{%
  \gdef\@bstract{%
   \parbox{\textwidth-0pc}{%
   \centerline{\bf Abstract}\penalty1000
   \noindent
   \renewcommand\baselinestretch{1.0}
   {#1}}}
}

\gdef\@email{}
\def\email#1{%
   \gdef\@email{%
   Email: {\tt #1}}
}

\def\ps@paper{\let\@mkboth\@gobbletwo%
     \ifnum\draftcontrol=1
        \def\@oddfoot{\hbox to \textwidth{\tiny \versionno \hfil\tiny\draftdate}%
        \hskip -\textwidth \hbox to \textwidth{\hfil\rm\thepage\hfil}}%
     \else\def\@oddfoot{\hbox to \textwidth{\hfil\rm\thepage\hfil}}
     \fi
     \let\@evenfoot\@oddfoot
}

\def\body{\clearpage
          \pagestyle{paper}
        }

\def\@version#1{\ifnum\draftcontrol=1
\typeout{}\typeout{#1}\typeout{}
\vskip3mm\centerline{\hbox{\fbox{\normalsize{\tt DRAFT -- #1 -- }
                   {\draftdate}}}}\vskip3mm
\fi}
\let\version\@version
\long\def\eqlabel#1{\ifnum\draftcontrol=1
                    \tag@false  
                    \tag*{(\theequation) \hbox to -0.2cm{\hspace{0cm}\small{#1}\hss}}
                    \refstepcounter{equation}
                    \edef\@currentlabel{\theequation}
                    \ltx@label{#1}
                    \else
                    \label{#1}
                    \fi
                    }
\let\st@bibitem\@bibitem
\let\st@lbibitem\@lbibitem
\ifnum\draftcontrol=1
  \def\@bibitem#1{%
    \st@bibitem{#1}\a@@label{#1}\ignorespaces}
  \def\@lbibitem[#1]#2{%
    \st@lbibitem[#1]{#2}\a@@label{#2}\ignorespaces}
  \def\a@@label#1{%
    \gdef\a@lab{\smash{\normalfont\small#1}}
    \ifvmode
      \if@inlabel
        \global\setbox\@labels\hbox{%
          \llap{\a@lab\let\a@lab\relax
                \kern\@totalleftmargin\kern\marginparsep}%
          \box\@labels}%
      \fi
    \fi}
\fi

\documentclass[12pt,letterpaper]{article}
\usepackage{amsmath}
\usepackage{amsmath}
\usepackage{amsmath}
\usepackage{amssymb}
\usepackage{amssymb}
\usepackage{amssymb}
\usepackage{amsmath,bm,amsfonts,amssymb,array,calc,amsthm,rotating,cite}
\usepackage{epsfig,psfrag}
\usepackage{graphicx}
\usepackage{color}
\usepackage[colorlinks=true]{hyperref}
\usepackage[all]{xy}

\renewcommand\baselinestretch{1.25}
\renewcommand\section{\@startsection {section}{1}{\z@}%
                                   {-3.5ex \@plus -1ex \@minus -.2ex}%
                                   {2.3ex \@plus.2ex}%
                                   {\normalfont\large\bfseries}}
\renewcommand\subsection{\@startsection{subsection}{2}{\z@}%
                                   {-3.25ex\@plus -1ex \@minus -.2ex}%
                                   {1.5ex \@plus .2ex}%
                                   {\normalfont\normalsize\bfseries}}
\renewcommand\subsubsection{\@startsection{subsubsection}{3}{\z@}%
                                   {-3.25ex\@plus -1ex \@minus -.2ex}%
                                   {1.5ex \@plus .2ex}%
                                   {\normalfont\normalsize\it}}
\renewcommand\paragraph{\@startsection{paragraph}{4}{\z@}%
                                   {-3.25ex\@plus -1ex \@minus -.2ex}%
                                   {1.5ex \@plus .2ex}%
                                   {\normalfont\normalsize\bf}}
\renewcommand\subparagraph{\@startsection{subparagraph}{5}{\z@}%
                                   {-1.25ex\@plus -1ex \@minus -.2ex}%
                                   {0ex \@plus .2ex}%
                                   {\normalfont\normalsize\it}}

\numberwithin{equation}{section}

\renewcommand*\l@section[2]{%
  \ifnum \c@tocdepth >\z@
    \addpenalty\@secpenalty
    \addvspace{.5em \@plus\p@}%
    \setlength\@tempdima{1.5em}%
    \begingroup
      \parindent \z@ \rightskip \@pnumwidth
      \parfillskip -\@pnumwidth
      \leavevmode \bfseries
      \advance\leftskip\@tempdima
      \hskip -\leftskip
      #1\nobreak\hfil \nobreak\hb@xt@\@pnumwidth{\hss #2}\par
    \endgroup
  \fi}
\renewcommand*\l@subsection{\addvspace{.0em \@plus\p@}\@dottedtocline{2}{1.5em}{2.3em}}
\renewcommand*\l@subsubsection{\addvspace{-.2em \@plus\p@}\@dottedtocline{3}{3.8em}{3.2em}}

\definecolor{refcol}{rgb}{0.0,0.0,0.2}
\definecolor{eqcol}{rgb}{.2,0,0}
\definecolor{purple}{cmyk}{0,1,0,0}

\gdef\@citecolor{refcol} \gdef\@linkcolor{eqcol}
\gdef\@urlcolor{refcol}
\def\colorlinkspurple{\gdef\@urlcolor{purple}}
\def\colorlinksblue{\gdef\@urlcolor{blue}}
\def\colorlinksred{\gdef\@urlcolor{red}}

\def\revise#1       {\raisebox{-0em}{\rule{3pt}{1em}}%
                     \marginpar{\raisebox{.5em}{\vrule width3pt\
                     \vrule width0pt height 0pt depth0.5em
                     \hbox to 0cm{\hspace{0cm}{%
                     \parbox[t]{4em}{\raggedright\footnotesize{#1}}}\hss}}}}

\def\sst{\scriptstyle}

\catcode`\@=12
\newcommand{\bqa}{\begin{eqnarray}}
\newcommand{\eqa}{\end{eqnarray}}
\begin{document}

\title{
Off-shell D-Brane/F-Theory Effective Superpotentials and Ooguri-Vafa Invariants of Several Compact Calabi-Yau Manifolds}

\author{
Shi Cheng,~~Feng-Jun Xu,~~Fu-Zhong Yang\footnote{Corresponding author~~~
E-mail: fzyang@ucas.ac.cn} \\[0.2cm]
\it College of Physical Sciences, University of Chinese Academy of Sciences\\
\it   YuQuan Road 19A, Beijing 100049, China}

\Abstract{~~~~According to the generalized GKZ system for CY manifold  with brane and the open-closed Mirror map, We calculate the off-shell D-Brane/F-Theory effective superpotentials and Ooguri-Vafa invariants of several compact Calabi-Yau manifolds . }

\makepapertitle

\body

\version\versionno

\vskip 1em

\newpage

\section{Introduction}
~~~~Superpotentials in the low energy limit can be early calculated by the method of mirror symmetry. In the low energy limit, superpotential encodes quantum correction to classical superpotential in IIA string theory. When Calabi-Yau manifold warped by several branes, there is a bit of complication to calculate open instanton correction. The calculation of superpotential have been well studied by the method of localization and mirror symmetry. Mirror symmetry as an effective method gives an equivalent between A model and B model and the later has the property of no quantum correction and can be calculated from geometry. Such that the equivalent relation, mirror map between A model moduli parameters and B model moduli parameters, is crucial. On the B model side, superpotential as a special solution of the differential system which derived form Picard-Fuch equations. In the more general situation that D-branes warp on the CY manifold and mirror to a divisor insecting on base CY manifold. Such that the natural mathematical interpretation of the B model is relative homology. This system equivalent to relative cohomology. For homological reason and relations between differential forms this system turns out to be a mixed Hodge structure. Which completely describe the differential system of superpotential. An useful method deals with the differential system is generalized GKZ system. Differential operators which annihilate the solutions of differential system are determined by the charge vectors as well as other operators got by cohomology's equivalent relation up to an exact form in type III case. In our paper we calculated two simple examples and didn't touch the complicated case for its no apparent relation to interpret direct integrality method.

The open superpotential encode the Ooguri-Vafa invariants which is the integral coefficients of power series of superpotential. The previous paper used the matrix factorization method to calculate the open string instanton number. On-shell superpotential was calculated as a special sector of off-shell superpotential. Previous paper got the on-shell superpotential as the substraction of off-shell superpotential of two holomorphic cycle which within the insertion of CY manifold and divisor. The substraction being the domain wall tension between these two cycles. After choosing the open moduli parameter describing the position of cycle within the complete insertion to be critical value the on-shell superpotential got. We just calculate the off-shell superpotential from integrating the special solutions of subsystem being the on-shell superpotential. On another words, the on-shell superpotential is obvious got by set the open moduli parameters to be critical values. The selection of mori cone which spanned by the charge vectors is quite crucial at different limit for the integral property of Ooguri-Vafa invariant. In our example we choose the same additional charge vector in \cite{Alim:2010za} which's exist due to the divisor. The result is ont integral because of unappropriate chosen of additional charge vector. After a maximal triangulation the result is again integral, but the on-shell superpotential is not easy to got in this method. We prefer the former unintegral form superpotential. The appropriate superpotential enjoy both advantages of former and later form is leave to our next paper\cite{Cheng}.

In section two we state the mathematical structure of off-shell superpotnetial and the extracting of instantons numbers form superpotnetial. In section three we illustrate the generalized GKZ system and correspondence Toric geometry and construction of CY hupersurface. In section four we calculate two examples and extract their open instantons numbers. In preparing the first three section we are inspired by \cite{Klevers:0811.2996,Klevers:1011.6375,Klevers:1106.6259v1}.

\section{Generalized GKZ system, off-shell superpotential and Ooguri-Vafa invariants}
\subsection{Mathematical structure of off-shell superpoential and Ooguri-Vafa invariants}

~~~~Type IIA and IIB string theory are $N=2$ supersymmetric theories. Mirror symmetry as a T-duality reveal the equivalence of the two theories. When D6-brane warps the special Lagrangian submanifold within CY manifold\cite{Aganagic:2000gs,Aganagic:2001nx}, the A brane within the submanifold is defined by
\bqa
\sum_i l_i^a |x_i|^2=c_a,~~~~\sum_i v_b^i\theta^i=0,~~~~\sum_i l_i^a v_b^i=0
\eqa
 where $ a,b=h^{1,1}(X)+1,h^{1,1}(X)+2,...,h^{1,1}(X)+N $. $N$ being the number of additional point in enhanced polyhedron. The later two equations are the condition of CY manifold. $c_a$ parameterize brane's position.

 The mirror B brane defined by the holomorphic divisor is D5-brane, the definition of which encode the additional charge vector.
 \bqa
 \prod_i y_i^{l_i^a}-\hat{z}_a=0,~~~~\hat{z}_a=\epsilon_a e^{-c_a}
 \eqa
 $ a=h^{2,1}(X^\star)+1,h^{2,1}(X^\star)+2,...,h^{2,1}(X^\star)+N$.

Our interest is the non-perturbative effect of superpotential being instantion correction to the classical superpotential which's classical part can be calculated from string theory. While its instantion correction can be calculated by topological string theory due to topological effective of inner space. For the D6-brane situation the correct action, Chern-Simons action, describes the brane action being brane superpotential \cite{Witten:1992fb}.
$$ \mathcal{W}_{A-brane}=\int_{X}\Omega^{3,0}\wedge \text{Tr}[A\wedge \bar{\partial}A+\frac{2}{3}A\wedge A\wedge A] $$
On the B model side, the brane superpotential being the following expression after a dimensional reduction because of D5-brane's warping a two cycle\cite{Jockers:2008pe}.
$$ \mathcal{W}_{B-brane}=\int_{\mathcal{C}}\Omega_{ijk}\phi^i\bar{\partial}_{z}\phi_j dzd\bar{z}$$
$\phi^i$ being the section of field base on two cycle $\mathcal{C}$.

The background $RR$ flux also contribute to the total superpotential.
$$\mathcal{W}_{flux}(z)\ =\int_{X} F^{(3)}\wedge  \Omega^{3,0}(z) $$

The two parts of B-model superpotential is classical, which can be calculated though geometry method. Although the natural mathematical structure of mirror manifold is the derive category, We will to state the complicated and beautiful interpretation in\cite{Cheng}. The relative cohomology is more simple but enough to calculate the superpotential. From the expression above, the brane superpotential can be expressed as the linear combination of two forms and flux contribution is the linear combination of three form. The parameters defining the deformations of cycles was encoded in the the brane superpotential, and the flux contribution obviously only involve the moduli parameters of base manifold\cite{Alim:2009bx}.
\bqa
\mathcal{W}_{brane}(z)\ =\int_{X}
\Omega^3(z,\hat{z}) = \sum_{\gamma_\alpha} \hat{N}_\alpha\hat{\Pi}^\alpha(z,\hat{z})\,
 ~~ N_\alpha \in \mathbb{Z}\,~~
  \gamma_\alpha \in H^3(z,\hat{z})~~\partial\gamma_\alpha\neq 0.
  \\
  \mathcal{W}_{flux}(z)\ =\int_{X}
F^{(3)}\wedge  \Omega^{3,0}(z) = \sum_{\gamma_\alpha} N_\alpha\Pi^\alpha(z)\ ,
 \qquad N_\alpha \in \mathbb{Z}\,~~~~
 \gamma_\alpha \in H^3(z).~~~~
\\
\mathcal{W}(z,\hat{z})=\sum_{\gamma_\alpha \in H^3(X^\star,E)}\underline{N}_\alpha\underline{\Pi}^\alpha(z,\hat{z}),~~~~\underline{N}_\alpha=(N_\alpha,\hat{N}_\alpha)~~~~~~~~~
~~~~~~~~~~~~~~~~~~~~~~~
\eqa

The classical geometry of B model with submanifold warped by D5-brane can be constructed within $N=1$ special geometry. Mixed Hodge structure which are powerful tools to calculate effective superpotential and derive Picard-Fuch equations in the language of relative cohomology of $(X^\star,E)$ \cite{Lerche:2002ck,Lerche:2002yw}.

The $2$ cycle $\mathcal{C}$ in $X^\star$ wrapped by brane carry brane charge. The physical string field is described by the fiber bundle based on the open-close moduli space $M$, such that its correspond object is cohomology of $M$. Since the relative cohomology form related to period integral, the fiber of $M$ is $H^3(X^\star,E)$. Define Hodge filtration $F^p=\oplus_qH^{q,3-p}$.
For any three form in the $H^3(X^\star,E)$,  ~$\underline{\Omega}$ can be split into 3-form $\Omega$ and colsed 2-form $\omega$ , satisfy the equivalent relation
$$
\int_{\underline{\Gamma}}\underline{\Omega}=\int_{\Gamma}\Omega+ \int_{ \partial \Omega} \omega , ~~~~  \underline{\Gamma} \in H_3(X^\star,E,\mathbb{Z})
$$

Apparently, $\omega$ is two form on submanifold $E$, such that if we split $H^{p,q}(X^\star,E)=H^{p,q}(X^\star)\oplus H^{p,q}(E)$ there exist the following mixed Hodge structure

$$
 \xymatrix{ F^3\cap H^3(X^\star)\ar[r]^{\sst
\partial_z}\ar[dr]^{\sst \partial_{\hat{z}}}& F^2\cap H^3(X^\star)\ar[r]^{\sst
\partial_z}\ar[dr]^{\sst \partial_{\hat{z}}}& F^1\cap H^3(X^\star)\ar[r]^{\sst
\partial_z}\ar[dr]^{\sst \partial_{\hat{z}}}&
F^0\cap H^3(X^\star)\ar[dr]^{\sst \partial_z,\partial_{\hat{z}}} \\
&F^2\cap H^3(X^\star,E)\ar[r]^{\sst \partial_z,\partial_{\hat{z}}}&
F^1\cap H^3(X^\star,E)\ar[r]^{\sst \partial_z,\partial_{\hat{z}}}&
F^0\cap H^3(X^\star,E)\ar[r]^{\sst \partial_z,\partial_{\hat{z}}}& 0}
$$

Any differential form relating to others under the action of variation of moduli parameter. The second line is the subsystem of the complete insertion. We can also derive the Picard-Fuchs equations of the $N=1$ special geometry from relations between these relative forms. The period integrals are the subset of solutions of these Picard-Fuch equations.

F-theory and Type IIA string theory are related to each other by circle compactifications from four to three\cite{Alim:2009bx,Alim:2010za}. Such that the relative periods of the brane compactification $(X^\star,E)$ can be identified as the holomorphic $(4,0)$ form on the fourfold $X^\star$ defined as the mirror moduli space of F-theory after compactficating the non-compact base $\mathbb{C}$ of $X^\star$ as $\mathbb{P}^1$ and $E$ being the fiber of $X^\star$ based on $\mathbb{C}$. At large base limit $X$ in A model can be mirror to the moduli space $X^\star_{c}$ the F-theory compactificated on. The open-close mirror symmetry was turn to close mirror symmetry in fourfold compactification of F-theory after these construction. The large volume limit correspond to weak coupling limit $g_s\rightarrow 0 $ under mirror symmetry. The Gukov-Vafa-Witten superpotentials $\mathcal{W}$ of F-theory in A model can be view as $g_s$ corrected $\mathcal{W}$ in large radius limit.$$\mathcal{W}_{GVW}=\mathcal{W}+\mathcal{O}(g_s)+\mathcal{O}(e^{-1/g_s}) $$ The method of our paper is to calculate the open-close superpotential $\mathcal{W}$ using the process of calculating $\mathcal{W}_{GVW}$, since$\mathcal{W}$ equals to $\mathcal{W}_{GVW}$ in large complex limit.

From the superpotential in the B model we can got the A model superpotential after changing of parameters because of the equality of A/B superpotential under mirror map between respective moduli parameters.
In A-model interpretation, the superpotential expressed in term of flat coordinates $(t,\hat{t})$, which relates to complex
structure parameters $(z,\hat{z})$ in B-model through mirror map, is
the generating function of the Ooguri-Vafa invariants
\cite{9912123}
\bqa
\mathcal{W}(t,\hat{t})=\sum_{\overrightarrow{k},\overrightarrow{m}}G_{\overrightarrow{k},\overrightarrow{m}}q^{d\overrightarrow{k}}\hat{q}^{d\overrightarrow{m}}=\sum_{\overrightarrow{k},\overrightarrow{m}}\sum_{d}n_{\overrightarrow{k},\overrightarrow{m}}\frac{q^{d\overrightarrow{k}}\hat{q}^{d\overrightarrow{m}}}{k^2}
\eqa where $q=e^{2\pi it}$, $\hat{q}=e^{2\pi i\hat{t}}$ and
$n_{\overrightarrow{k},\overrightarrow{m}}$ are Ooguri-Vafa
invariants counting disc instantons in relative
homology class $(\overrightarrow{m},\overrightarrow{k})$, where
$\overrightarrow{m}$ represents the elements of $H_1(D)$ and
$\overrightarrow{k}$ represents an element of $H_2(X)$.
$G_{\overrightarrow{k},\overrightarrow{m}}$ are open Gromov-Witten
invariants. From string world-sheet
viewpoint, these terms in the superpotential represents the
contribution from instantons of sphere and disk.

\subsection{Generalized GKZ system, submanifold and type II/F off-shell superpotentials}
~~~~The equal footing of CY geometry and brane geometry in GLSM make toric geometry be possible to describe A/B model with brane by standard toric method. The mirror pair $(X,X^\star)$ being a pair of hupersurface in two different toric ambient space $W$ and $W^\star$, $W=P_\Sigma(\Delta)$, $W^\star=P_\Sigma(\Delta^\star)$, $(\Delta,\Delta^\star)$ being reflexive polyhedron pairs, which defined as the convex of hull of $p$ integral vertices, including integral points except vertices being on the surfaces and edges of polyhedron. $P_\Sigma(\Delta)$ is the toric variety defined by $\Delta$. The zero locus of $P_\Sigma(\Delta)$ is the CY three fold $X$ we consider, such that polyhedrons $\Delta^\star$ are defined in four dimensions and can be viewed as embodying in a hupersurface $H$ in five dimensions\cite{Hosono:1993qy,Hosono:1995bm}.

    $p$ integral point $\nu_i^\star $ of $\Delta^\star$ correspond to $p$ homogeneous coordinates $x_i$ on $W$ and satisfy $h^{1,1}(X)$ linear relations $\sum_i l_i^a \nu_i=0$. Vector $l^a$reprent the $U(1)$ charge of GLSM on $X$,  the torus transformation of coordinates is
    \bqa x_i\rightarrow \lambda_a^{l_i^a}x_i, ~~ \lambda_a \in \mathbb{C}^\star ,~~ a=1,...,h^{1,1}(X) \eqa

    For a weight projective space $X_d(1,\omega_2,\omega_3,\omega_4,\omega_5)$, the vertices of $\Delta$ are the following
\bqa \begin{split}
    &\nu_1=(1,1,1,1),~~~&\nu_2=(1-d/\omega_2,1,1,1,1),~~&\nu_3=(1,1-d/\omega_3,1,1),\\
    &\nu_4=(1,1,1-d/\omega_4,1),~~&\nu_5=(1,1,1,1-d/\omega_5),~~ ~~&d=\sum_i\omega_i+1.
    \end{split} \eqa
The integral points in dual $\Delta^\star$ include both vertices and other integral points on face or edge of $\Delta^\star$, the vertices of $\Delta^\star$ are
\bqa
\begin{split}
&\nu_1^\star=(\omega_2,\omega_3,\omega_4,\omega_5),~~~&\nu_2^\star=(-1,0,0,0),~~~&\nu_3^\star=(0,-1,0,0),\\
&\nu_4^\star=(0,0,-1,0),                              ~~~&\nu_5^\star=(0,0,0,-1),~~~&\nu_0^\star=(0,0,0,0).
\end{split}
\eqa
Other integral points on the $\Delta^\star$ are early got from the liner combination of these five vertices.

    The map between integral points and monomials is
    \bqa
P=\sum_{\nu_i^\star \in \Delta^\star }a_i X^{\nu_i^\star},~~~~ X^{\nu_i^\star}=\prod_{k=1}^{4}X_k^{\nu_{i,k}^\star},~~~~\nu_i^\star=( \nu_{i,1}^\star,\nu_{i,2}^\star,\nu_{i,3}^\star,\nu_{i,4}^\star)
\eqa
$\nu^\star$ being the vertices and some other integral points on $\Delta^\star$, $a_i$ being complex parameters. In principle all integral points on the face and edge of $\Delta^\star$ should contribute to the expression of hupersurface $P$, but the deformation of polynomials determined by some integral points are not indepent to each others. After reduction by chance $a_i$, only part of integral points on the face and edge are selected to determine $P$, and the number or deformation point is equal to the number of Kailer moduli of $X$. Since our analysis is in the weight projective space, we apply the homogeneous coordinates $x_j$ defined on weighted projective space, $(2.10)$ can be rewritten as
\bqa
P=\sum_{i=0}^{p-1}a_i\prod_{\nu_j \in \Delta}x_j^{<\nu_j,v_i^\star>+1}
\eqa

    The zero locus of $P$ is the CY hupersurface of the close string interal spacetime. Since our CY surface with one brane, we should extend analysis to relative cohomology. The $(\Delta,\Delta^\star)$ should be extended to $(\overline{\Delta},\overline{\Delta}^\star)$, added one vertices outside the hupersurface $H$ the $\Delta$ located at. The additional one vertices correspondence to one additional $a_i$, which correspondence to the open moduli parameter.

    In fact, the two-cycles wrapped by branes are only holomorphic cycles with respect to the variation of complex structure of CY mirror manifold, but the requirement condition is that the moduli parameters must be located at the critical points of superpotentials. For off-shell superpotentail this requirement guaranteed only at special values of open moduli parameters. However according to statement of  \cite{Klevers:0811.2996,Klevers:1011.6375,Klevers:1106.6259v1}, the non-holomorphic two-cycles can be replaced by holomorphic divisor of ambient CY space with the divisor $D$ encompassing the two-cycles.

   The open sector for D-brane being described by a family of hupersurface as $(2.10)$ which is compatible with the definition of B-brane in section one.
   \bqa
   Q(D)=\sum_{i=p}^{p+p'-1}a_i X_k^{\bar\nu_{i,k}^\star}
   \eqa
   The divisor can be got from integral points on the $\Delta^\star$. The above definition of divisor is the same as hupersurface $P$ and we usually choose the vertices points as the definition of divisor.
   $Q(D)$ can also be written in homogeneous coordinates the same like $(2.11)$. The CY surface with brane is the submanifold being the complete insertion of divisor $Q(D)$ and base manifold  $P=0$.
   \bqa
   P_D=P|_{Q(D)}
   \eqa
    $l_i^a$ which define mori cone extended to $\underline{l_i^a}$, satisfy
    \bqa
    \sum_{i=0}^{p-1}\underline{l_i^a}=0,~~~~\sum_{i=p}^{p+p'-1}\underline{l_i^a}=0,~~~~\sum_{i=0}^{p+p'+1}\underline{l_i^a}\bar{\nu_i}^\star=0
    \eqa
$p'$ is the number of additional vertices. The dimension of $\nu_i$ extended from four to five. In the case of adding just a few points, the usually construction of extended vertices is ggyhghy
\bqa
 \bar{\nu_i}^{\star}=\left\{
\begin{aligned}
(\nu_i^{\star},0)~~~~~i=0,1,...,p-1 \\
(\nu_i^{\star},1)~~i=p,...,p+p'-1
\end{aligned}\right.
\eqa

    $\overline{\Delta}^\star$ live in five dimensions describe the four fold in toric geometry. Since II/F-theory compactificated on CY fourfold $X_4^\star$, which can be associated to $\overline{\Delta}^\star$. That is the nontrivial idea to get the II/F theory superpotential from extended GKZ system of relative cohomology of open-close B model at tree level.

    On the side of B model, the fundamental period is the integral of homogenous three form $\Omega$, which are chosen as \cite{Jockers:2008pe}
    \bqa
    \Pi_\alpha=\int_{\Gamma^\alpha} \frac{\text{log} Q(\hat{z})}{P(z)}\prod_{k=1}^{4}\frac{d X_k}{X_k},~~~~\Gamma^\alpha \in H_3(X^\star,E,\mathbb{Z})
    \eqa

    A differential system can be constructed makes these period integral be the solutions of a series of differential operators. The differential operator describes the symmetry of these periods. Fortunately, these symmetry encoded by $\underline{l}^a$. Due to the insertion of a divisor $D$ on $X$, the description of symmetries of complete insertion is modified by adding more charge vector $\underline{l}^a$. We need to add one additional $\underline{l}^a$ when one brane situation. Hence we need not use the complicated form $(2.16)$ to calculation, just need to return to the typical form
    \bqa
    \Pi_\alpha=\int_{\Gamma^\alpha} \frac{1}{P(z)}\prod_{k=1}^{4}\frac{d X_k}{X_k},~~~~\Gamma^\alpha \in H_3(X^\star,E,\mathbb{Z})
    \eqa
    Differential operators of the above fundamental periods are
    \bqa
     \eqlabel{operator}
   \begin{split}
   &\mathcal{L}(l^a)=\prod_{l_i^a>0}(\partial_{a_i})^{l_i^a}-\prod_{l_i^a<0}(\partial_{a_i})^{l_i^a}\\
   &\mathcal{Z}_k=\sum_{i=0}^{p-1}\nu_{i,k}^{\star}\vartheta_i,
   \qquad \mathcal{Z}_0=\sum_{i=0}^{p-1}\vartheta_i-1
   \end{split}
    \eqa
    The charge $l^a$ in the above equations is $\underline{l}^a$ for simplification. Where $\vartheta_i = a_i\partial_{a_i}$. The above operators annihilate periods: $\mathcal{L}(l^a)\Pi=0$ and $\mathcal{Z}_k\Pi=0$. $\mathcal{Z}$ describes
    the gauge transformation $(2.7)$, which is torus action. Such that there exist an equivalent relation: $\Pi(a_i)\sim\Pi(z)$. After comparing with $(2.11)$ and defining the torus invariant algebraic coordinates $z_a=(-)^{l_0^a}\prod_i a_i^{l_i^a}, a=1,2...h^{2,1}(X^\star)$, $\vartheta_i$ can be written as linear combination of $\partial_{z_a}$ by partial derivation, $\vartheta_i=\sum_a l_i^a$. Then operators $\mathcal{L}(l^a)$ was rewritten as
    \bqa
    \mathcal{L}(l^a)=\prod_{k=1}^{l_0}(\vartheta_0-k)\cdot \prod_{l_i^a>0}\prod_{k=0}^{l_i^a-1}(\vartheta_i+k)-(-1)^{l_0^a}z_a\prod_{k=1}^{-l_0^a}(\vartheta_0-k)\cdot\prod_{l_i^a<0}\prod_{k=0}^{-l_i^a-1}(\vartheta_i-k).~~~~
    \eqa

    If the charge vector $l^a$ are appropriately chosen by linear combination of $l^a$, the generating function of solutions of GKZ system is
    \bqa
    B_{l^a}(z_a;\rho_a)=\sum_{n_1,...,n_N\in
\mathbb{Z}_0^{+}}\frac{\Gamma(1-\sum_a
l_0^a(n_a+\rho_a))}{\prod_{i>0}\Gamma(1+\sum_a
l_i^a(n_a+\rho_a))}\prod_a z_a^{n_a+\rho_a}.
    \eqa
    Which is also the same expression for generalized GKZ system. The restriction of GKZ system of $P=0$ to the submanifold $P_D$ is the subsystem which's integration being off-shell superpotential.

In this paper we choose two vertices of $\Delta^\star$ to define $Q(D)$, then the open moduli parameters is one in projective space. The expression of it in the homogeneous coordinates is
  \bqa
   Q(D)=x_1^{b_1}+\hat{z}x_2^{b_2}
  \eqa
   where $b_1,~b_2$ are appropriate integers which is same as their correspondence $x_i$ terms in $P$. The relative 3-form $\underline{\Omega}:=(\Omega_{X^\star}^{i,3-i},0)$ and the relative
periods satisfy a set of differential equations\cite{Alim:2010za}
 \bqa
\mathcal{L}_a(\theta,\hat{\theta})\underline{\Omega}=d\underline{\omega}^{(2,0)}~\Rightarrow~\mathcal{L}_a(\theta,\hat{\theta})\mathcal{T}(z,\hat{z})=0.
\eqa
with some corresponding two-form $\underline{\omega}^{(2,0)}$ which's exist due to equivalent relation of relative cohomology. The domain wall tension $\mathcal{T}(z,\hat{z})$ is the substraction of two off-shell superpoential vanishing on two different cycle respectively. In our paper the two cycles are chosen to be involution with a $\mathbb{Z}_2 $ symmetry, such that the expression of two off-shell superpotentials got by direct integrality are equal to each other if we don't care the sign of $\hat z$. The differential operators $\mathcal{L}_a(\theta,\hat{\theta})$ can
be expressed as
\bqa
\mathcal{L}_a(\theta,\hat{\theta}):=\mathcal{L}_a^{blk}-\mathcal{L}_a^{bdy}\hat{\theta}~\Rightarrow~\mathcal{L}_a^{blk}\mathcal{T}(z,\hat{z})=\mathcal{L}_a^{bdy}\mathcal{T}(z,\hat{z})
\eqa
for $\mathcal{L}_a^{blk}$ acting only on bulk part from closed
sector, the first line of mixed Hodge structure, which is the same when the mirror CY manifold without brane. $\mathcal{L}_a^{bdy}$ act on boundary part from open-closed
sector, the second line of mixed Hodge structure, and $\hat{\theta}=\hat{z}\partial_{\hat{z}}$. The explicit
form of these operators will be given in following model. From the
$(2.22)$ one can obtain
\bqa \eqlabel{sub} 2\pi
i\hat{\theta}\mathcal{T}(z,\hat{z})=\pi(z,\hat{z})
\eqa
$\pi(z,\hat{z}) $ is the period of two forms of the submanifold. The critical locus for calculating the on-shell superpotential is the area on which this two form period equals zeros.

The special solution of subsystem is\cite{Alim:2010za}
\bqa
\pi(u_1,u_2,...)=\frac{1}{2}B_{\tilde{l}}(u_1,u_2,...;\rho_1,\rho_2,...)
\eqa
in which $\tilde{l_i}$ being the charge vectors of subsystem and $\rho_i$ being the index of differential subsystem. $\tilde{l}$ can be derived from the algebraic expression of complete intersection by $(2.13)$.
The critical locus off-shell superpotentials are independent to open moduli parameters.
From $(2.22)$ and $(2.23)$ one can obtain differential equation with the inhomogeneous
term $f_a(z)$ at the critical points \bqa
\mathcal{L}_a^{b}T(z)=f_a(z) \eqa and \bqa 2\pi i
f_a(z)=\mathcal{L}_a^{bd}\pi(z,\hat{z})|_{\hat{z}=\text{critical
points}}
 \eqa
\subsection{The calculation of mirror map}
 For calculation of instanton corrections, one need to know mirror map. The closed-string periods are \cite{Hosono:1993qy}
  \begin{equation} \label{Wp}
  \Pi(z) = \left(\begin{array}{c}
                    \omega_0(z,\rho)|_{\rho=0}\\
                    D_i^{(1)} \omega_0(z,\rho)|_{\rho=0}\\
                    D_i^{(2)} \omega_0(z,\rho)|_{\rho=0}\\
                    D^{(3)}   \omega_0(z,\rho)|_{\rho=0}\\
                  \end{array}
            \right).
\end{equation}
where $i=1,...,h_{21}(X^\ast)$,
  \bqa w_0=\sum c(n_i+\rho_i)z^{n_i+\rho_i} \qquad i=1,2,3.
 \eqa
 \bqa
c(n_i+\rho_i)=\frac{\Gamma(\Sigma_{k=1}^{3}l_0^k(n_k+\rho_k)+1)}{\Pi_{i=1}^a\Gamma(\Sigma_{k=1}^3l_i^k(n_k+\rho_k)+1)}
 \eqa
 and \begin{equation}
  D_i^{(1)} := \partial_{\rho_i}, ~ D_i^{(2)} := \frac{1}{2} \kappa_{ijk} \partial_{\rho_j} \partial_{\rho_k},
  ~ D^{(3)} := -\frac{1}{6} \kappa_{ijk} \partial_{\rho_i} \partial_{\rho_j}
  \partial_{\rho_k}
\end{equation}
 $\kappa_{ijk}$ is intersection number of X.

     The flat coordinates in A-model at large radius regime are related to the flat coordinates of B-model at large complex structure regime by mirror map $t_i=\frac{\omega_i}{\omega_0},~\omega_i:= D_i^{(1)}
 \omega_0(z,\rho)|_{\rho=0}$.
  When CY manifold with brane the open mirror map is similar as above and the only difference is the adding of charge vector of divisor. When the charge vectors are mixed, the close mirror maps will turn out to be the combination of correspondence $\omega_i$ for keeping the unmixing of $z_i$\cite{Alim:2011rp}.

\section{Open Ooguri-Vafa invariants of several examples}

\subsection{Hypersurface $X_{12}(1,2,2,3,4)$ }
 ~~~~The GKZ system of $X_{14}(1,3,3,3,5)$ have been calculated in \cite{Alim:2010za}. We cite the charge vectors
  \begin{equation}
\begin{tabular}{c|c c c c c c c c c c}
~  & $0$ & $1$& $2$& $3 $ & $4$ & $5$ & $6$ &$7$ &$8$   \\\hline
$l^1$ & $-6$ & $-1$ & $1$& $1$ & $0$ & $2$ & $3$ & $0$ & $0$\\
$l^2$ & $0$ & $1$& $0$  & $0$& $1$ & $0$ & $-2$ & $0$ & $0$ \\
$l^3$ & $0$ &$0$& $1$ & $-1$ & $0$ & $0$ & $0$ & $-1$  & $1$
\end{tabular}
\end{equation}
 The charge vector $l^3$ is determined by the divisor we chosen.
The charge vectors of subsystem are
\begin{equation}
\begin{tabular}{c|c c c c c c c c}
~  & $0$ & $1$& $2$& $3 $ & $4$ & $5$\\\hline
$\tilde{l}^1$ & $-6$ & $-1$ & $2$ & $0$ & $2$ & $3$\\
$\tilde{l}^2$ & $0$ & $1$& $0$& $1$ & $0$ & $-2$
\end{tabular}
\end{equation}

 The off-shell superpotential being the integration of special periods of subsystem \cite{Alim:2010za}
 \bqa
 \pi(u_1,u_2)=\frac{1}{2}B_{\tilde{l}}(u_1,u_2;\frac{1}{2},0),~~u_1=-\frac{z_1}{z_3}(1-z_3)^2,~~u_2=z_2
  \eqa
 in which $z_3$ is the open moduli parameter in divisor.
\bqa
\begin{split}
&\mathcal{W}(z_1,z_2,z_3)=\frac{1}{2\pi i}\int_{i}^{\sqrt{z_3}}\pi(u_1,u_2) \frac{d\sqrt{z_3}}{\sqrt{z_3}}\\
&=\sum_{n_1,n_2\geq0}\frac{\Gamma(4+6n_1)\frac{2i}{\pi}}{\Gamma(2+2n_1)^2 \Gamma(1+n_2)\Gamma(1/2-n_1+n_2)\Gamma(5/2+3n_1-2n_2)}\cdot\\
&~~~~\frac{1}{2}( -_2F_1(-1-2n_1,-\frac{1}{2}-n_1,\frac{1}{2}-n_1,z_3+1)z_2^{n_2}(-)^{n_1+\frac{1}{2}}(\frac{z_1}{z_3+1})^{\frac{1}{2}+n_1}+\\
&~~~~\frac{4^{\frac{1}{2}+n_1}\sqrt{\pi}\Gamma(\frac{1}{2}-n_1)z_1^{\frac{1}{2}+n_1}z_2^{n_2}}{\Gamma(-n_1)} )
\end{split}
\eqa

The inverse mirror maps in terms of $q_i=e^{2\pi i t_i}$ got from the method of last section are
\bqa
\begin{split}
& z_1=q_1+60q_1^2-1530q_1^3-1530q_1^3+3q_1q_2-2232q_1^2q_2+3q_1q_2^2-q_1q_3\\&~~~~~~-90q_1^2q_3-3q_1q_2q_3+...\\
& z_2=q_2-60q_1q_2+5130q_1^2q_2-2q_2^2-300q_1q_2^2+3q_2^3+30q_1q_2q_3+... \\
& z_3=q_3+q_3^2-90q_1q_2q_3^2-120q_1q_2q_3^3+...
\end{split}
\eqa
Using the modified multi-cover formula for this case
  \bqa
\frac{\mathcal{W}}{w_0}=\frac{1}{(2 \pi i)^2}\sum_{k,~ d_{1,2,3}~ odd\geq
0}n_{d_1,d_2,d_3}\frac{q_1^{kd_1/2}q_2^{kd_2}q_3^{kd_3}}{k^2}
\eqa the superpotentials $\mathcal{W}$ give Ooguri-Vafa invariants $n_{d_1,d_2,d_3}$ for the
normalization constants
  $c=1$.~Some result are listed in the table 1. We don't list the result of $d_3=0$, being the chose instanton numbers, for its equal to the on-shell result in \cite{Alim:2010za}.

 \begin{table}[!h]
\def\temptablewidth{1.0\textwidth}
\begin{center}
\begin{tabular*}{\temptablewidth}{@{\extracolsep{\fill}}c|ccccccc}
$d_3=1$&&&&&&&\\
$d_1\diagdown d_2 $&$0$     &$1$          &$2$      &$3$      &$4$          \\\hline
$1$                     & $-8 $     &$-24$                            &$0$                       &$ 0 $                 &$0$          \\
  $3$                     & $\frac{2656}{3}$     &$9648$                         &$-116064 $    &$-52112$            & $0$             \\
   $5$                      &$\frac{-777608}{5}$      &$1057912$                   &$12665072$              &$-451067392 $              & $-713479128$            \\
   $7$                        & $\frac{236871872}{7}$      &$\frac{-1192932144}{5}$              &$\frac{8818538496}{5}$               &$33198114384$      & $-2080499160192$    \\
   $9$                          & $\frac{-519563127296}{63}$ &$\frac{3204825296040}{49}$              &$\frac{-117752914909008}{245}$    &$\frac{173218640866272}{35}$  &$\frac{-597187778407632}{5}$
    \end{tabular*}
       {\rule{\temptablewidth}{1pt}}
\end{center}
 \def\temptablewidth{1.0\textwidth}
\begin{center}
\begin{tabular*}{\temptablewidth}{@{\extracolsep{\fill}}c|ccccccc}
$d_3=2$&&&&&&&\\
$d_1\diagdown d_2$ &$0$             &$1$               &$2$                 &$3$                &$4$          \\\hline
$1$             & $0$               &$0$               &$0$                  &$ 0 $               &$0$          \\
  $3$          & $\frac{-664}{3}$     &$2112 $          &$27816 $            &$12128$            & $0$             \\
   $5$         &$\frac{727056}{5}$      &$-1703968 $     &$26472368$              &$546373488$              & $843670752 $            \\
   $7$       & $\frac{-372224472}{7}$   &$561417120$       &$-4924251912 $               &$103752174400$      & $4115818961304$    \\
   $9$      & $\frac{162953910032}{9}$ &$-198743023488 $     &$\frac{8198797933776}{5}$    &$\frac{-88997168551728}{5}$  &$\frac{2589936048140976}{5}$
    \end{tabular*}
       {\rule{\temptablewidth}{1pt}}
\end{center}
\def\temptablewidth{1.0\textwidth}
\begin{center}
\begin{tabular*}{\temptablewidth}{@{\extracolsep{\fill}}c|ccccccc}
$d_3=3$&&&&&&&\\
$d_1\diagdown d_2$ &$0$             &$1$               &$2$                 &$3$                &$4$          \\\hline
$1$             & $0$               &$0$               &$0$                  &$ 0 $               &$0$          \\
  $3$          & $\frac{-512}{9}$     &$1176  $          &$9312 $            &$\frac{14680}{3}$            & $0$             \\
   $5$         &$\frac{-131376}{5}$      &$444112 $     &$-6903632$              &$118575312$              & $-177301248 $            \\
   $7$       & $\frac{222220336}{7}$   &$-447008520$       &$4642741392 $               &$-96786651368$      & $-2998245394560$    \\
   $9$      & $\frac{-493088587712}{9}$ &$256376134464 $     &$-2410157075664$    &$27146117195600$  &$794487818407152$
    \end{tabular*}
      {\rule{\temptablewidth}{1pt}}
       \tabcolsep 0pt \caption{Ooguri-Vafa invariants $n_{d_1,d_2,d_3}$ for the off-shell
superpotential $\mathcal{W}$ on the 3-fold $\mathbb{P}_{1,2,2,3,4}[12]$} \vspace*{-12pt}
\end{center}
       \end{table}
\newpage

        It is unfortunate that this chosen of charge vectors is inappropriate, because once all nonzero integral number in any column of these three $l$ have same sign the fractional result is unavoidable. One can remedy the result by making a maximal triangulation of $l$ to make the open instanton numbers we calculated to be integral. Since our target is to use the direct integrality method to get the off-shell superpotential and once we made such a triangulation the on-shell superpotential will become incorrect in large complex limit, such that  we proper to choose these charge vectors. The expression of off-shell superpotential sames complicated, and changing the charge vector will remedy this fault. To find a good expression of off-shell superpotential and from which after reduction we can get on-shell superpotential and its corresponding open and close integral topological invariants is leave to\cite{Cheng} . The above off-shell superpotential can not guarantee the integral of instanton numbers, but up to an  total constant number we can choose the special value of open moduli parameters, equals one in our example, to got the close instanton numbers which is integral.

\subsection{Hypersurface $X_{14}(1,2,2,2,7)$}
~~~~The charge vectors of $X_{14}(1,2,2,2,7)$ are
  \begin{equation}
\begin{tabular}{c|c c c c c c c c c c}
~  & $0$ & $1$& $2$& $3 $ & $4$ & $5$ & $6$ &$7$ &$8$  \\\hline
$l^1$ & $-7$ & $-3$ & $1$& $1$ & $1$ & $0$ & $7$ & $0$ & $0$ \\
$l^2$ & $0$ & $1$& $0$  & $0$& $0$ & $1$ & $-2$ & $0$ & $0$ \\
$l^3$ & $0$ &$0$& $1$ & $-1$ & $0$ & $0$ & $0$ & $-1$ & $1$
\end{tabular}
\end{equation}
 The charge vector $l^3$ is determined by the divisor we chosen. The charge vectors of subsystem are
\begin{equation}
\begin{tabular}{c|c c c c c c c c c c}
~  & $0$ & $1$& $2$& $3 $ & $4$ & $5$   \\\hline
$\tilde{l}^1$ & $-7$ & $-3$ & $1$& $2$ & $0$ & $7$  \\
$\tilde{l}^2$ & $0$ & $1$& $0$  & $0$ & $1$ & $-2$
\end{tabular}
\end{equation}
 Special period of subsystem is
   \bqa
 \pi(u_1,u_2)=\frac{1}{2}B_{\tilde{l}}(u_1,u_2;\frac{1}{2},0),~~u_1=-\frac{z_1}{z_3}(1-z_3)^2,~~u_2=z_2
  \eqa
 in which $z_3$ is the open moduli parameter in divisor. The off-shell superpotenatial as the integration of special period of subsystem
\bqa
\begin{split}
&\mathcal{W}(z_1,z_2,z_3)=\frac{1}{2\pi i}\int_{i}^{\sqrt{z_3}}\pi(u_1,u_2) \frac{d\sqrt{z_3}}{\sqrt{z_3}}\\
&=\sum_{n_1,n_2\geq0}\frac{\Gamma(\frac{9}{2}+7n_1)\frac{2i}{\pi}}{\Gamma(\frac{3}{2}+n_1)\Gamma(2+2n_2)\Gamma(\frac{9}{2}+7n_1-2n_2)\Gamma(1+n_2)\Gamma(-\frac{1}{2}-3n_1+n_2)}\cdot\\
&~~~~\frac{1}{1+2n_1}(-_2F_1(-1-2n_1,-\frac{1}{2}-n_1,\frac{1}{2}-n_1,z_3+1)z_2^{n_2}(-)^{n_1+\frac{1}{2}}(\frac{z_1}{z_3+1})^{\frac{1}{2}+n_1}+\\
&~~~~\frac{4^{\frac{1}{2}+n_1}\Gamma(\frac{1}{2}-n_1)\sqrt{\pi}z_1^{\frac{1}{2}+n_1}z_2^{n_2}}{\Gamma(-n_1)})
\end{split}
\eqa

The inverse mirror maps in terms of $q_i=e^{2\pi i t_i(z_1,z_2,z_3)}$ are
\bqa
\begin{split}
& z_1=q_1+6q_1^2+9q_1^3+7q_1q_2-56q_1^2q_2+21q_1q_2^2-q_1q_3-9q_1^2q_3-7q_1q_2q_3+...\\
& z_2=q_2-2q_1q_2+5q_1^2q_2-2q_2^2+36q_1q_2^2+3q_2^3+q_1q_2q_3+... \\
& z_3=q_3+q_3^2-210q_1q_2^3q_3^2+q_3^3+q_3^4+q_3^5+...
\end{split}
\eqa
Some open Ooguri-Vafa invariants are listed in table 2.
 \begin{table}[!h]
\def\temptablewidth{1.0\textwidth}
\begin{center}
\begin{tabular*}{\temptablewidth}{@{\extracolsep{\fill}}c|ccccccc}
$d_3=1$&&&&&&&\\
$d_1\diagdown d_2     $&$0$                 &$1$              &$2$                    &$3$      &$4$          \\\hline
$1$                     & $\frac{1}{2} $     &$-7$               &$\frac{-35}{2}$        &$ 0 $                     &$0$          \\
  $3$                     & $\frac{-17}{6}$     &$49$                &$\frac{-777}{2}$     &$\frac{-6209}{3}$           & $\frac{-55167}{2}$             \\
   $5$                      &$\frac{113}{5}$      &$-574$                &$\frac{12999}{2}$  &$-43883$                     & $\frac{461335}{2}$            \\
   $7$                        & $\frac{-8548}{35}$   &$\frac{41566}{5}$    &$\frac{-1297793}{10}$ &$\frac{18593974}{15}$      & $\frac{254937907}{30}$    \\
   $9$                          & $\frac{960448}{315}$ &$\frac{-651881}{5}$   &$\frac{26028593}{10}$ &$\frac{-3397403831}{105}$  &$\frac{60181826993}{210}$
    \end{tabular*}
       {\rule{\temptablewidth}{1pt}}
\end{center}
   \def\temptablewidth{1.0\textwidth}
\begin{center}
\begin{tabular*}{\temptablewidth}{@{\extracolsep{\fill}}c|ccccccc}
$d_3=2$&&&&&&&\\
$d_1\diagdown d_2$ &$0$             &$1$                   &$2$                  &$3$                     &$4$          \\\hline
$1$             & $0$               &$0$                   &$0$                  &$ 0 $                   &$0$          \\
  $3$          & $\frac{17}{24}$     &$\frac{-49}{4}$      &$\frac{777}{8} $     &$\frac{-1421}{3}$         & $\frac{50617}{8}$             \\
   $5$         &$\frac{-1053}{40}$    &$686 $              &$\frac{67473}{8}$    &$\frac{391097}{6}$              & $\frac{-4729529}{12} $            \\
   $7$       & $\frac{25671}{56}$   &$\frac{-63175}{4}$    &$\frac{508641}{2} $  &$\frac{-10183299}{4}$      & $\frac{220870531}{12}$    \\
   $9$      & $\frac{-2830931}{360}$ &$\frac{6785373}{20}$ &$-6908076$           &$\frac{2653125727}{30}$  &$\frac{-97462036973}{120}$
    \end{tabular*}
       {\rule{\temptablewidth}{1pt}}
\end{center}
   \def\temptablewidth{1.0\textwidth}
\begin{center}
\begin{tabular*}{\temptablewidth}{@{\extracolsep{\fill}}c|ccccccc}
$d_3=3$&&&&&&&\\
$d_1\diagdown d_2$ &$0$                 &$1$                  &$2$                     &$3$                      &$4$          \\\hline
  $1$           & $0$                   &$0$                  &$0$                     &$ 0 $                    &$0$          \\
  $3$          & $\frac{35}{144} $      &$\frac{-35}{8} $     &$\frac{665}{16} $       &$\frac{-9415}{36} $      & $\frac{45605}{16}$             \\
   $5$         &$\frac{213}{40}$        &$\frac{-567}{4} $    &$\frac{14833}{8}$       &$\frac{-378539}{24}$     & $\frac{2491237}{24}$            \\
   $7$       & $\frac{-8929}{28}$       &$11137$              &$\frac{-2966523}{16} $  &$\frac{3901989}{2}$      & $\frac{-239497041}{16}$    \\
   $9$      & $\frac{3955121}{432}$     &$\frac{-3189501}{8}$ &$\frac{132502125}{16} $ &$\frac{-3929352707}{36}$ &$\frac{4153932475}{4}$
    \end{tabular*}
       {\rule{\temptablewidth}{1pt}}
       \tabcolsep 0pt \caption{Ooguri-Vafa invariants $n_{d_1,d_2,d_3}$ for the off-shell
superpotential $\mathcal{W}$ on the 3-fold $\mathbb{P}_{1,2,2,2,7}[14]$} \vspace*{-12pt}
\end{center}
       \end{table}
\newpage

\subsection{Hypersurface $X_{18}(1,1,1,6,9)$}
~~~~The charge vectors of  $X_{18}(1,1,1,6,9)$ are
  \begin{equation}
\begin{tabular}{c|c c c c c c c c c c}
~  & $0$ & $1$& $2$& $3 $ & $4$ & $5$ & $6$ &$7$ &$8$  \\\hline
$l^1$ & $-6$ & $0$ & $0$& $0$ & $2$ & $3$ & $1$ & $0$ & $0$ \\
$l^2$ & $0$ & $1$& $1$  & $1$& $0$ & $0$ & $-3$ & $0$ & $0$ \\
$l^3$ & $0$ &$1$& $-1$ & $0$ & $0$ & $0$ & $0$ & $-1$ & $1$
\end{tabular}
\end{equation}
 The charge vector $l^3$ is determined by the divisor we chosen.The charge vectors of subsystem are
    \begin{equation}
\begin{tabular}{c|c c c c c c c c c}
~  & $0$ & $1$& $2$& $3 $ & $4$ & $5$  \\\hline
$\tilde{l}^1$ & $-6$ & $0$ & $0$& $2$ & $3$ & $1$  \\
$\tilde{l}^2$ & $0$ & $1$  & $2$& $0$ & $0$ & $-3$
\end{tabular}
\end{equation}
 Special period of subsystem is
  \bqa
 \pi(u_1,u_2)=\frac{1}{2}B_{\tilde{l}}(u_1,u_2;0,\frac{1}{2}),~~u_1=z_1,~~u_2=-\frac{z_2}{z_3}(1-z_3)^2
  \eqa
 in which $z_3$ is the open moduli parameter in divisor. The off-shell superpotenatial as the integration of special period of subsystem
\bqa
\begin{split}
&\mathcal{W}(z_1,z_2,z_3)=\frac{1}{2\pi i}\int_{i}^{\sqrt{z_3}}\pi(u_1,u_2) \frac{d\sqrt{z_3}}{\sqrt{z_3}}\\
&=\sum_{n_1,n_2\geq0}\frac{\Gamma(1+6n_1)\frac{2i}{\pi}}{\Gamma(1+2n_1)\Gamma(1+3n_1)\Gamma(-\frac{1}{2}+n_1-3n_2)\Gamma(\frac{3}{2}+n_2)\Gamma(2+2n_2)}\cdot\\
&~~~~\frac{1}{1+2n_2}(-_2F_1(-1-2n_2,-\frac{1}{2}-n_2,\frac{1}{2}-n_2,z_3+1)z_1^{n_1}(-)^{n_2+\frac{1}{2}}(\frac{z_2}{z_3+1})^{\frac{1}{2}+n_2}+\\
&~~~~\frac{4^{\frac{1}{2}+n_1}\sqrt{\pi}\Gamma(\frac{1}{2}-n_2)z_2^{\frac{1}{2}+n_2}z_2^{n_2}}{\Gamma(-n_2)})
\end{split}
\eqa

The inverse mirror maps in terms of $q_i=e^{2\pi i t_i(z_1,z_2,z_3)}$  are
\bqa
\begin{split}
& z_1=q_1-312q_1^2+87084q_1^3-2q_1q_2+1788q_1^2q_2+5q_1q_2^2+q_1q_2q_3+...\\
& z_2=q_2-180q_1q_2+20790q_1^2q_2+9q_2^3-q_2q_3+180q_1q_2q_3+... \\
& z_3=q_3+q_3^2+q_3^3+q_3^4+q_3^5-6126120q_1^3q_2q_3^2...
\end{split}
\eqa
Some open instanton numbers are listed in table 3.
 \begin{table}[!h]
\def\temptablewidth{1.0\textwidth}
\begin{center}
\begin{tabular*}{\temptablewidth}{@{\extracolsep{\fill}}c|ccccccc}
$d_3=1$&&&&&&&\\
$d_2\diagdown d_1     $&$0$                 &$1$              &$2$                    &$3$      &$4$          \\\hline
$1$                     & $-\frac{1}{2} $     &$135$               &$\frac{35235}{2}$        &$ 564555 $                     &$9812556$          \\
  $3$                     & $\frac{17}{6}$     &$-945$                &$\frac{319545}{2}$     &$-29151855 $           & $\frac{65914264485}{2}$             \\
   $5$                      &$\frac{-113}{5}$      &$11070 $                &$-\frac{4851495}{2}$  &$341382115$                     & $-\frac{174105836235}{2}$            \\
   $7$                        & $\frac{8548}{35}$   &$-160326$    &$\frac{96752637}{2}$ &$-\frac{4572171963}{5}$      & $\frac{3619051217541}{2}$    \\
   $9$                          & $-\frac{960448}{315}$ &$\frac{17600787}{7}$   &$-\frac{13589269803}{14}$ &$\frac{1654946034516}{7}$  &$-\frac{686548806079149}{14}$
    \end{tabular*}
       {\rule{\temptablewidth}{1pt}}
\end{center}
   \def\temptablewidth{1.0\textwidth}
\begin{center}
\begin{tabular*}{\temptablewidth}{@{\extracolsep{\fill}}c|ccccccc}
$d_3=2$&&&&&&&\\
$d_2\diagdown d_1$ &$0$             &$1$                   &$2$                  &$3$                     &$4$          \\\hline
$1$             & $0$               &$0$                   &$0$                  &$ 0 $                   &$0$          \\
  $3$          & $\frac{-17}{24}$     &$\frac{945}{4}$      &$-\frac{319545}{8} $     &$\frac{24046755}{4}$         & $\frac{64535887485}{8}$             \\
   $5$         &$\frac{1053}{40}$    &$-13230$              &$\frac{27010665}{8}$    &$-604609030$              & $\frac{573479123985}{4} $    \\
   $7$       & $\frac{25671}{56}$   &$\frac{1218375}{4}$    &$-\frac{194476545}{2} $  &$\frac{40090855671}{2}$      & $-\frac{16154915828679}{4}$    \\
   $9$      & $\frac{2830931}{360}$ &$\frac{26172153}{4}$ &$2613075471$           &$-\frac{2684085142773}{4}$  &$\frac{1157187494566647}{8}$
    \end{tabular*}
       {\rule{\temptablewidth}{1pt}}
\end{center}
   \def\temptablewidth{1.0\textwidth}
\begin{center}
\begin{tabular*}{\temptablewidth}{@{\extracolsep{\fill}}c|ccccccc}
$d_3=3$&&&&&&&\\
$d_1\diagdown d_2$ &$0$                 &$1$                  &$2$                     &$3$                      &$4$          \\\hline
  $1$           & $0$                   &$0$                  &$0$                     &$ 0 $                    &$0$          \\
  $3$          & $-\frac{35}{144} $      &$\frac{675}{8} $     &$-\frac{319545}{16} $       &$\frac{29260785}{8} $      & $-\frac{45405899325}{16}$             \\
   $5$         &$-\frac{213}{40}$        &$\frac{10935}{4} $    &$-\frac{6161265}{8}$       &$\frac{649575985}{4}$     & $-\frac{147665561085}{4 }$            \\
   $7$       & $\frac{8929}{28}$       &$-214785$              &$\frac{1160612955}{16} $  &$-\frac{130432131753}{8}$      & $\frac{54973444140603}{16}$    \\
   $9$      & $-\frac{3955121}{432}$     &$\frac{61511805}{8}$ &$-\frac{50772734685}{16} $ &$\frac{6854310497595}{8}$ &$-\frac{1538427241683783}{8}$
    \end{tabular*}
       {\rule{\temptablewidth}{1pt}}
       \tabcolsep 0pt \caption{Open Ooguri-Vafa invariants $n_{d_1,d_2,d_3}$ for branes on $\mathbb{P}_{1,1,1,6,9}[14]$} \vspace*{-12pt}
\end{center}
       \end{table}

\newpage

\subsection{Hypersurface $X_{18}(1,2,3,3,9)$}
~~~~The charge of  $X_{18}(1,2,3,3,9)$ are
  \begin{equation}
\begin{tabular}{c|c c c c c c c c c c c}
~  & $0$ & $1$& $2$& $3 $ & $4$ & $5$ & $6$ &$7$ &$8$ &$9$ \\\hline
$l^1$ & $-6$ & $-1$ & $0$& $1$ & $1$ & $3$ & $2$ & $0$ &$0$& $0$ \\
$l^2$ & $0$ & $1$& $0$  & $0$& $0$ & $0$ & $-2$ &$1$& $0$ & $0$ \\
$l^3$ & $0$ &$0$& $1$& $0$ & $0$ & $0$ & $1$ & $-2$ & $0$ & $0$\\
$l^4$ & $0$ &$0$& $0$& $1$ & $-1$ & $0$ & $0$ & $0$ & $-1$ & $1$\\
\end{tabular}
\end{equation}
 The charge vector $l^4$ is determined by the divisor we chosen. The charge vectors of subsystem are
   \begin{equation}
\begin{tabular}{c|c c c c c c c c c c}
~  & $0$ & $1$& $2$& $3 $ & $4$ & $5$ & $6$  \\\hline
$\tilde{l}^1$ & $-6$ & $-1$ & $0$& $2$ & $3$ & $2$ & $0$ \\
$\tilde{l}^2$ & $0$ & $1$& $0$  & $0$ & $0$ & $-2$ &$1$ \\
$\tilde{l}^3$ & $0$ &$0$& $1$& $0$ & $0$ & $1$ & $-2$ \\
\end{tabular}
\end{equation}
 Special period of subsystem is
  \bqa
 \pi(u_1,u_2,u_3)=\frac{1}{2}B_{\tilde{l}}(u_1,u_2,u_3;\frac{1}{2},0,0),~~u_1=-\frac{z_1}{z_4}(1-z_4)^2,~~u_2=z_2,~~z_3=z_3
  \eqa
 in which $z_4$ is the open moduli parameter in divisor. The off-shell superpotenatial as the integration of special period of subsystem
\bqa
\begin{split}
&\mathcal{W}(z_1,z_2,z_3,z_4)=\frac{1}{2\pi i}\int_{i}^{\sqrt{z_4}}\pi(u_1,u_2) \frac{d\sqrt{z_4}}{\sqrt{z_4}}\\
&=\sum_{n_1,n_2,n_3\geq0}\frac{\Gamma(4+6n_1)\frac{2i}{\pi}}{\Gamma(2+2n_1)\Gamma(\frac{5}{2}+3n_2)\Gamma(\frac{1}{2}-n_1+n_2)\Gamma(1+n_3)\Gamma(2+2n_1-2n_2+n_3)}\cdot\\
&~~~~\frac{1}{1+2n_1}( -_2F_1(-1-2n_1,-\frac{1}{2}-n_1,\frac{1}{2}-n_1,z_4+1)z_2^{n_2}z_3^{n_3}(-)^{n_1+\frac{1}{2}}(\frac{z_1}{z_4+1})^{\frac{1}{2}+n_1}+\\
&~~~~\frac{4^{\frac{1}{2}+n_1}\sqrt{\pi}\Gamma(\frac{1}{2}-n_1)z_1^{\frac{1}{2}+n_1}z_2^{n_2}z_3^{n_3}}{\Gamma(-n_1)} )
\end{split}
\eqa

The inverse mirror maps in terms of $q_i=e^{2\pi i t_i(z_1,z_2,z_3,z_4)}$ got from the method of last section are
\bqa
\begin{split}
& z_1=q_1+60q_1^2-1530q_1^3+2q_1q_3-744q_1^2q_2+q_1q_2^2+2q_1q_2q_3-q_1q_4-90q_1^2-2q_1q_2q_4+...\\
& z_2=q_2-60q_1q_2+5130q_1^2 q_2-2q_2^2+240q_1q_2^2+3q_2^3+q_2q_3-60q_1q_2q_3-3q_2^2q_3+... \\
& z_3=q_3+q_2q_3-300q_1q_2q_3+5130q_1^2q_2q_3-300q_1q_2^2q_3-2q_3^2-3q_2q_3^2+1380q_1q_2q_3^2+...\\
& z_4=q_4+q_4^2-30q_1q_2q_4^2-30q_1q_2q_3q_4^2+q_4^3+40q_1q_2q_4^3+...
\end{split}
\eqa
Some open instanton numbers are listed in table 4.
 \begin{table}[!h]
\def\temptablewidth{1.0\textwidth}
\begin{center}
   {\rule{\temptablewidth}{1pt}}
\begin{tabular*}{\temptablewidth}{@{\extracolsep}c|c|c|c|c|ccc}
   $(d_1,d_2,d_3,d_4)$  &$n_{d_1, d_2, d_3, d_4}$  &$(d_1, d_2, d_3, d_4)$  &$n_{d_1, d_2, d_3, d_4}$  &$(d_1, d_2, d_3, d_4)$  &$n_{d_1, d_2, d_3, d_4}$         \\\hline
   $(1,0,0,1)$          & $-\frac{1}{2} $          &$(3,3,1,1)$             &$-201$                    &$(5,3,0,2)$             &$89352 $       \\
   $(3,0,0,1)$          & $\frac{166}{3}$          &$(5,3,1,1)$             &$2923869$                 &$(3,1,1,2)$             &$44 $        \\
   $(5,0,0,1)$          &$-\frac{97201}{10}$       &$(3,2,2,1)$             &$-201$                    &$(5,1,1,2)$             &$\frac{106498}{3} $            \\
   $(1,1,0,1)$          & $-\frac{1}{2}$           &$(5,2,2,1)$             &$-12755$                  &$(3,2,1,2)$             &$\frac{1101}{2} $    \\
   $(3,1,0,1)$          & $-201$                   &$(3,3,2,1)$             &$-201$                    &$(5,2,1,2)$             &$\frac{1412477}{3}$            \\
   $(5,1,0,1)$          &$\frac{132239}{6}$        &$(5,3,2,1)$             &$-2923869$                &$(3,3,1,2)$             &$44 $       \\
   $(3,2,0,1)$          &$-201$                    &$(3,3,3,1)$             &$\frac{166}{3}$           &$(5,3,1,2)$             &$3553224 $                   \\
   $(5,2,0,1)$          &$-12755$                  &$(5,3,3,1)$             &$-12755$                  &$(3,2,2,2)$             &$44 $      \\
   $(3,3,0,1)$          &$\frac{166}{3}$           &$(3,0,0,2)$             &$\frac{83}{6}$            &$(5,2,2,2)$             &$89352 $          \\
   $(5,3,0,1)$          &$12755$                   &$(5,0,0,2)$             &$\frac{45441}{5}$         &$(3,3,2,2)$             &$44 $         \\
   $(1,1,1,1)$          &$-\frac{1}{2}$            &$(3,1,0,2)$             &$44$                      &$(5,3,2,2)$             &$3553224 $            \\
   $(3,1,1,1)$          &$-201$                    &$(5,1,0,2)$             &$-\frac{106498}{3}$       &$(3,3,3,2)$             &$-\frac{83}{6} $           \\
   $(5,1,1,1)$          &$\frac{132239}{6}$        &$(3,2,0,2)$             &$44$                      &$(5,3,3,2)$             &$89352 $          \\
   $(3,2,1,1)$          &$-2277$                   &$(5,2,0,2)$             &$89352$                   &$$                                    \\
   $(5,2,1,1)$          &$-\frac{788582}{3}$       &$(3,3,0,2)$             &$-\frac{83}{6}$           &$$
    \end{tabular*}
       {\rule{\temptablewidth}{1pt}}
       \tabcolsep 0pt \caption{Ooguri-Vafa invariants $n_{d_1,d_2,d_3}$ for branes on $\mathbb{P}_{1,2,3,3,9}[14]$} \vspace*{-12pt}
\end{center}
       \end{table}

\newpage

\section{Summary }
~~~~Although open instanton numbers calculated in our method is not integral, we believe it's right. The physical means of instanton numbers is the number of BPS states which must be integral, such that we should remedy the fault of our result to turn open instanton numbers to be integral. We thought the combination of charge vectors is right way, but it might intrigue another problem that the close instanton number can not direct got from superpotential. Find a beautiful expression to extract both the close and open instanton numbers is quite meaningful.

The example of one open moduli parameters is quite special. The more general case is several open moduli parameters and several divisors and the base CY manifold can be complete insertion. Easy example is the case of several parallel divisors which correspond to several parallel branes.

Furthermore, to obtain the D-brane superpotential from the $\mathcal{A}_{\infty}$-structure of the derived category and the path algebras of quiver is crucial.

\end{document}